\documentclass[11pt,letterpaper]{article}
\usepackage[T1]{fontenc}
\usepackage[utf8]{inputenc}
\usepackage{lmodern}
\DeclareUnicodeCharacter{00B2}{\ensuremath{^2}}
\DeclareUnicodeCharacter{039B}{\ensuremath{\Lambda}}
\DeclareUnicodeCharacter{03B3}{\ensuremath{\gamma}}
\DeclareUnicodeCharacter{03BB}{\ensuremath{\lambda}}
\DeclareUnicodeCharacter{03C0}{\ensuremath{\pi}}
\DeclareUnicodeCharacter{03F1}{\ensuremath{\rho}}
\DeclareUnicodeCharacter{2013}{--}
\DeclareUnicodeCharacter{2019}{'}
\DeclareUnicodeCharacter{2212}{-}
\DeclareUnicodeCharacter{221E}{\ensuremath{\infty}}
\DeclareUnicodeCharacter{2264}{\ensuremath{\leq}}
\DeclareUnicodeCharacter{2265}{\ensuremath{\geq}}
\usepackage[margin=1in]{geometry}
\usepackage{amsmath,amssymb,amsthm}
\usepackage{microtype}
\usepackage{times}
\usepackage{booktabs,longtable,array,ragged2e,multirow,color}
\usepackage{graphicx,float,placeins}
\usepackage{caption}
\usepackage{subcaption}
\usepackage{enumitem}
\usepackage[round,authoryear]{natbib}
\usepackage[hidelinks]{hyperref}
\usepackage{url}
\title{\textbf{The Gold Rush in AI4Math: Where Are We Now?}}
\author{Jiashun Jin\footnote{jiashunmail@gmail.com} \and Zheng Tracy Ke\footnote{zke@fas.harvard.edu} \and Bingcheng Sui\footnote{bingchengsui@g.harvard.edu}}
\date{}
\begin{document}
\maketitle

\begin{abstract}
Recent advances in artificial intelligence (AI) have sparked growing interest
in its use for mathematical research. While some view this as a major
opportunity for discovery, others have raised concerns about its impact on
traditional research practices. Despite extensive debate, empirical evidence
on how AI is actually being used in mathematics remains limited.

To address this gap, we collected all 32,944 arXiv submissions posted between
March 1 and August 20, 2026, whose primary or secondary categories included
Mathematics. We identified 3,575 submissions that explicitly disclosed author
use of AI, of which 1,712 involved at least one substantive mathematical
contribution.

Our analysis reveals several broad patterns. First, disclosed AI use increased
sharply over the study period, with substantive use growing from 1.39\% of
Mathematics submissions in March to 14.09\% through August 20. Second,
substantive AI use is highly uneven across fields: Combinatorics has the
largest number of such papers, while Metric Geometry has the highest
substantive-use rate. Third, substantive AI use is geographically
concentrated: under weighted author counts, the United States and China
together account for about two-thirds of the recognized country weight.
Fourth, AI is already being applied to open research problems: among 717 named
open-problem records associated with substantive use, 71\% are labeled
as fully resolved based on the authors' descriptions, with proofs of the
conjectured statement more common than counterexamples or disproofs. Finally,
AI-system use is also highly concentrated, with OpenAI systems appearing most
frequently, followed by Anthropic.

Together, these findings suggest that AI-assisted mathematics is expanding
rapidly but remains at an early and uneven stage of adoption. They provide an
empirical picture of where AI is beginning to influence mathematical research
and a data-driven basis for ongoing discussions about its future role in the
discipline.
\end{abstract}

\section{Introduction}
%Mathematics has long served both as a testing ground for artificial
%intelligence and as a domain in which computational tools can augment human
%reasoning. The rapid development of large language models (LLMs) and other
%reasoning systems has substantially changed this relationship. AI systems are
%no longer limited to performing numerical computation or symbolic
%manipulation; increasingly, they can generate conjectures, construct multi-step
%arguments, search for proofs, and interact with formal proof assistants. A
%series of recent results illustrates the pace of this progress. FunSearch
%combined a large language model with an automated evaluator to discover
%improved constructions for the cap-set problem in extremal combinatorics
%\citep{romera2024funsearch}, while AlphaGeometry achieved near-Olympiad-level
%performance on challenging geometry problems using a neuro-symbolic approach
%\citep{trinh2024alphageometry}. More recently, AlphaProof combined
%reinforcement learning with the Lean proof assistant and, together with
%AlphaGeometry 2, achieved silver-medal-level performance at the 2024
%International Mathematical Olympiad \citep{hubert2026alphaproof}.

%The recent development of artificial intelligence (AI) has inspired growing interest in using AI to accelerate scientific discovery. 
Mathematics, one of the most logically rigorous
disciplines, has served as a natural test ground for evaluating the
reasoning capabilities of AI systems
\citep{hendrycks2021measuring,lewkowycz2022solving,trinh2024solving}.
Early efforts focused primarily on solving well-specified mathematical
problems, ranging from elementary exercises to competition-level problems.
Recent systems such as FunSearch, AlphaGeometry, and AlphaProof have pushed
this frontier further, demonstrating increasingly strong capabilities in
mathematical discovery, geometry, and formal reasoning
\citep{romera2024funsearch,trinh2024solving,hubert2026alphaproof}.
These advances raise a more ambitious question:
\emph{Can AI move beyond solving well-specified problems and contribute
meaningfully to research-level mathematics?}
%\citep{romeraparedes2024mathematical}

%The frontier has since begun to move from competition mathematics toward
%mathematical research itself. Recent systems have been designed not merely to
%solve problems with known solutions, but to explore conjectures and open
%research questions. For example, \citet{feng2026autonomous} report experiments
%in which an AI research agent was used on professional-level mathematical
%problems and open conjectures. In a separate large-scale study,
%\citet{tsoukalas2026formal} developed an LLM-based formal proof-search agent
%and evaluated it on hundreds of open problems, reporting proofs for several
%previously open Erd\H{o}s problems and dozens of conjectures from the Online
%Encyclopedia of Integer Sequences. These developments suggest a potentially
%important transition: AI is evolving from a tool for testing mathematical
%reasoning into a tool that may participate directly in the production of new
%mathematics.

There has been encouraging recent progress. In early 2026, Claude Opus 4.6 found a general construction for
an open Hamiltonian-cycle decomposition problem %posed by Donald Knuth
\citep{knuth2026claude}. Soon afterward, an unreleased OpenAI model produced a
counterexample to Erd\H{o}s's nearly 80-year-old unit-distance conjecture
\citep{alon2026unitdistance}. The \emph{First Proof} project offered a more
systematic test: on ten unpublished research problems, current AI systems
obtained passing solutions to seven under independent expert refereeing
\citep{abouzaid2026firstproof}. In July, Alp\"oge and Claude produced an
explicit counterexample to the Jacobian conjecture in dimension three
\citep{alpoge2026jacobian}. In August, OpenAI reported that its Astra model had
resolved or substantially advanced ten long-standing problems across several
areas of mathematics and theoretical computer science
\citep{openai2026tenadvances}. Around the same time, an unreleased version of
Claude improved the unconditional lower bound for the proportion of simple
zeros of the Riemann zeta function on the critical line from $41.6\%$ to
$67.2\%$ \citep{claude2026zeta}. Together, these examples suggest that AI is
beginning to contribute directly to the production of new mathematics.

This transition has generated both excitement and concern within the
mathematical community. Recent perspective articles have begun to examine how
AI may reshape mathematical practice, understanding, and professional norms
\citep{avigad2026mathematicians,klowden2026mathematical,
commelin2026shaping,tao2026mathematics}. On the one hand, AI systems may serve
as powerful research assistants by suggesting arguments, searching for
counterexamples, formalizing reasoning, and accelerating mathematical
exploration. Formal proof assistants such as Lean can also help verify
AI-generated proofs mechanically
\citep{song2025leancopilot,hubert2026alphaproof}. On the other hand, increasing
AI use raises concerns about reliability, attribution and credit, access to
proprietary models, and the role of human understanding and judgment
\citep{commelin2026shaping,tao2026mathematics}. These issues have become
sufficiently prominent that the 2026 \emph{Leiden Declaration on Artificial
Intelligence and Mathematics} called for transparency and for preserving
established norms of verification, attribution, openness, and human
responsibility \citep{alper2026leiden}.

%This transition has generated both considerable excitement and substantial
%concern within the mathematical community. On the one hand, AI systems may
%serve as powerful research assistants: they can explore large spaces of
%possible arguments, suggest intermediate lemmas, search for counterexamples,
%formalize informal reasoning, and potentially reduce the time required to test
%mathematical ideas. Formal proof assistants such as Lean also provide a
%mechanism for checking AI-generated proofs mechanically, thereby mitigating
%some of the well-known reliability problems of LLM-generated reasoning
%\citep{song2025leancopilot,hubert2026alphaproof}. On the other hand, the
%increasing role of AI raises questions about mathematical reliability,
%attribution and credit, access to proprietary models, the allocation of
%research attention, and the role of human understanding and judgment in
%mathematical discovery. These concerns have become sufficiently prominent that
%the 2026 \emph{Leiden Declaration on Artificial Intelligence and Mathematics},
%endorsed by the International Mathematical Union, called for greater
%transparency and for preserving established norms of verification, attribution,
%openness, and human responsibility in AI-assisted mathematics
%\citep{alper2026leiden}.

\begin{figure}[tb!]
\centering
\includegraphics[width=0.98\textwidth, trim=0 20 0 0, clip=true]{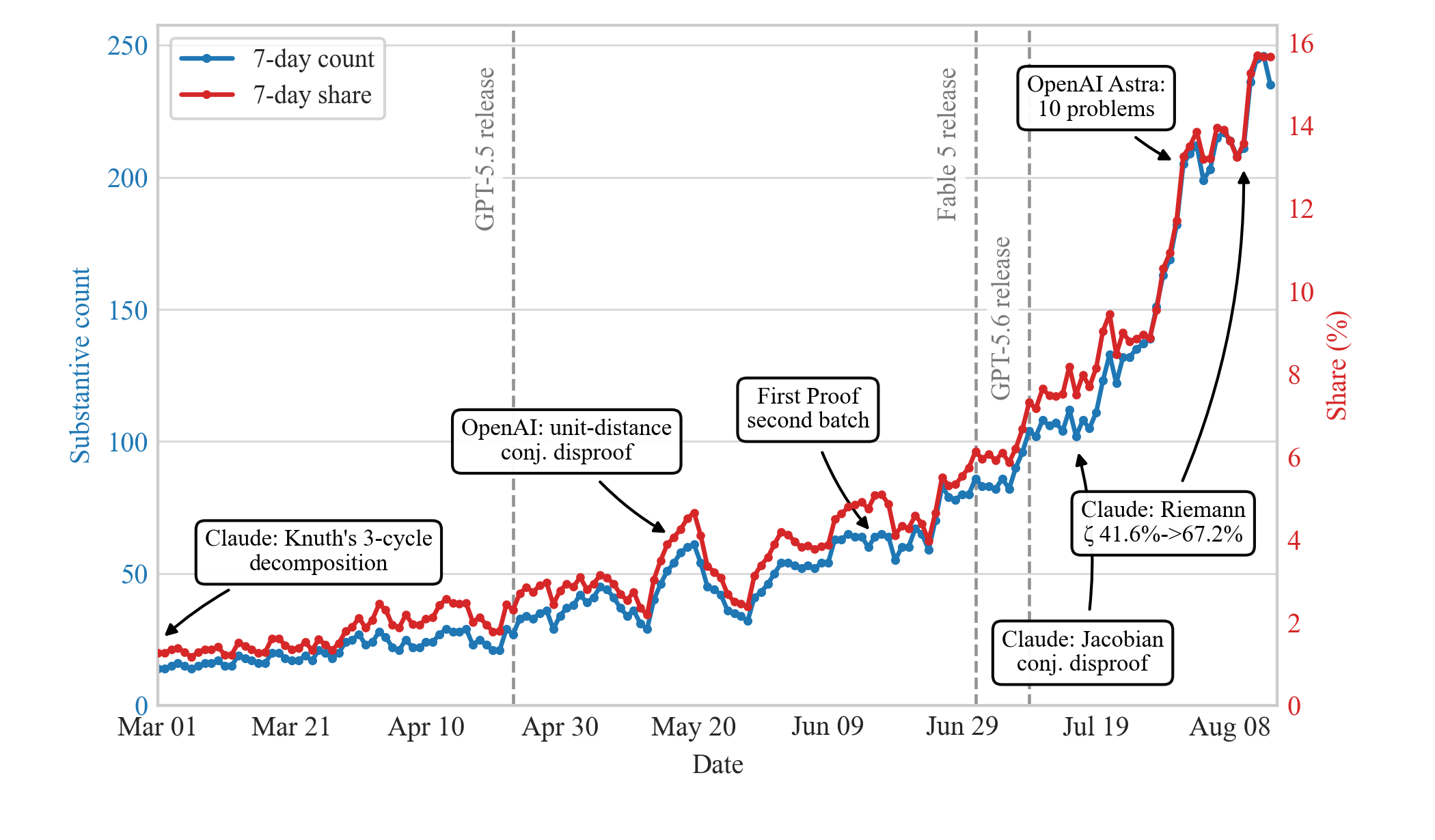}
\caption{Substantive AI use in seven-day forward windows. Each daily marker reports the
number of submissions with substantive AI use and their share among all
Mathematics submissions in the complete seven-day window beginning on that
date. Dashed vertical lines indicate model release dates, while rounded
callouts mark notable AI-assisted research events. Details of the data
construction are provided in Section~\ref{sec:data}.}
\label{fig:overall-7day-news}
\end{figure}

Despite the rapidly growing literature on the \emph{capabilities} of AI
systems in mathematics, much less is known about their \emph{actual adoption
in mathematical research}. Existing studies largely ask whether AI can generate formally verified proofs or tackle selected
research-level problems. These studies help characterize the technological
frontier, but they do not reveal how frequently mathematicians are using AI in
practice, how adoption is evolving over time, which areas of mathematics are
adopting these tools most rapidly, or which AI systems are most commonly used.
This distinction can be summarized by two different questions:
``Can AI do mathematical research?'' and ``To what extent is AI already being
used to do mathematical research?'' The latter is an empirical question for
which systematic evidence remains limited.

%Despite the rapidly growing literature on the \emph{capabilities} of AI systems
%in mathematics, considerably less is known about their \emph{actual adoption
%in mathematical research}. Most existing studies ask questions such as whether
%a model can solve Olympiad problems, generate formally verified proofs, or
%resolve selected research-level conjectures. These experiments are essential
%for understanding the technological frontier, but they do not tell us how
%frequently working mathematicians are using AI in research, whether adoption
%is increasing over time, which areas of mathematics are adopting these tools
%most rapidly, where the researchers using them are located, or which AI
%systems are most commonly employed. In other words, there is an important
%distinction between asking ``Can AI do mathematical research?'' and asking
%``To what extent is AI already being used to do mathematical research?'' The
%latter question is empirical, and systematic evidence remains limited.

To study actual AI adoption in mathematical research, we examine recent
submissions to arXiv. We collected all 32,944 submissions under the Mathematics category between March 1 and August 20, 2026, and identified
3,575 papers with confirmed disclosed author use of AI, including 1,712 with
at least one substantive mathematical contribution. Under our non-exclusive
coding, 1,225 papers report AI involvement in proof construction. Figure~\ref{fig:overall-7day-news} provides an
overview of the rapid growth of substantive AI use over this period. 
Using these data, we study how disclosed AI use has evolved over time, how it
varies across mathematical fields, what roles AI plays in research, its
involvement with named open problems, the authors and institutions adopting
these tools, the AI systems being used, and selected characteristics of
AI-assisted papers. Our reliance on explicit disclosure \citep{arxiv2026aipolicy} provides a transparent
and reproducible criterion for identifying AI-assisted research, but should be
interpreted as a conservative measure of adoption because undisclosed uses
cannot be observed. 

Our results reveal a rapidly expanding but highly uneven pattern of AI
adoption in mathematics. Disclosed AI use rises sharply over the study period,
with substantive use growing particularly quickly. Adoption is concentrated
in a small number of mathematical fields, led by Combinatorics and Number
Theory, and is also concentrated among a relatively small set of authors and
institutions. We find substantial AI involvement with named open problems,
including both unresolved questions and problems reported as resolved. Model
use is similarly concentrated, with OpenAI systems appearing most frequently followed by Anthropic. Taken together, these findings suggest that
AI-assisted mathematics is expanding rapidly, but remains at an early and
highly heterogeneous stage of adoption.

%Our results point to a rapidly expanding but highly uneven pattern of adoption.
%Disclosed AI use increases sharply over the study period, with substantive use
%growing even faster from a small base. Substantive-use papers are concentrated
%in a few mathematical areas, led by Combinatorics and Number Theory, while
%some fields show little or no substantive use. Authors using AI for proof
%construction are geographically concentrated, with the United States and China
%together accounting for approximately 18\% of the weighted author records. We also
%identify 717 named open-problem records associated with substantive AI use;
%a minority remain open or only partially resolved, while roughly 71\% are
%reported as fully resolved, with true resolutions substantially more common
%than false ones. Model use is similarly concentrated, with OpenAI systems
%appearing most frequently and Anthropic a distant second. Together, these
%patterns suggest that AI-assisted mathematics is expanding rapidly but remains
%at an early and uneven stage of adoption.

By providing a systematic snapshot of recent arXiv activity, our study
complements the rapidly expanding literature on AI's mathematical capabilities
with evidence on its adoption in research practice. We aim to establish
an empirical baseline for a debate that has so far been shaped largely by
technological demonstrations, individual case studies, and competing
predictions about the future of the field. As AI systems continue to improve,
systematically tracking how, where, and by whom they are used will be important
for understanding how the practice of mathematical research itself is
changing.

%By providing a systematic snapshot of recent arXiv activity, our study
%complements the rapidly expanding literature on AI's mathematical capabilities
%with evidence about its adoption in research practice. Rather than attempting
%to determine whether AI is beneficial or detrimental to mathematics, we seek
%to establish an empirical baseline for a debate that has so far been driven
%largely by technological demonstrations, individual case studies, and
%competing predictions about the future of the discipline. As AI systems
%continue to improve, tracking how, where, and by whom these tools are used may
%be essential for understanding how the practice of mathematical research
%itself is changing.

\section{Data and study design}\label{sec:data}
Between March 1 and August 20, 2026, there were 32,944 submissions whose
primary or secondary categories included Mathematics. We downloaded the PDF
files for all submissions, together with their arXiv identifiers, posting
dates, titles, category labels, and authors. Each PDF was
converted to plain text using advanced document-parsing tools. Specifically,
text was extracted page by page, with page boundaries explicitly marked. For
articles exceeding 300,000 characters, only the beginning and end 150,000 characters of the text
were retained. Among the 32,944 records, only 1,007 were flagged as truncated.

\medskip
\noindent 
{\bf The questionnaire.}
We designed a questionnaire to analyze each submission and collect information
that cannot be directly obtained from the arXiv metadata. The first group of
questions concerns the categories of AI use. We consider two main categories,
{\it substantive use} and {\it non-substantive use}, each consisting of four
subcategories; see Table~\ref{tab:ai-categories-combined}. For each
subcategory, the questionnaire requires a yes/no response, and the
subcategories are non-exclusive. The {\it substantive use} category is labeled
yes if at least one of its four subcategories is labeled yes; the same rule
applies to {\it non-substantive use}.

The second group of questions identifies whether the article addresses any
open problems or conjectures. For each such problem or conjecture, the
questionnaire records its status as {\it open}, {\it partially resolved},
{\it resolved true}, {\it resolved false}, or {\it unclear}, and whether AI
played a critical role in reaching the reported resolution.

The third group of questions concerns the authors' institutions and countries.
Because arXiv records provide author names but generally not institutional
affiliations, this information is obtained from the parsed full text---for
example, from listed affiliations or email addresses---or, when necessary,
through web searches.

\medskip
\noindent 
{\bf Collection of questionnaire responses.}
We used AI to answer the questionnaire for each submission. Importantly, the
responses are based on the authors' disclosures rather than on 
assessment of the mathematical content. Therefore, the AI system used for this
task does not need to perform mathematical reasoning. Its primary role is to
identify relevant disclosure statements about AI use and extract their
semantic meaning. 

We used GPT-5.6 Terra with low reasoning effort and processed
the submissions in parallel using 100 worker threads.
The prompt consisted of a fixed system instruction and an article-specific user
message. The system instruction defined the scope of AI, specified a three-level
confidence scale, required all supporting evidence to be quoted verbatim from
the article, and presented the questionnaire described above. The user message
provided the arXiv metadata and enclosed the extracted article text. 
The output was restricted to a single structured object with four components:
AI-use judgments, named AI models, qualifying open problems, and
author--institution pairs. For AI-use judgments, it requested confidence scores from
1 to 3 for each subcategory, 
%Binary AI-use judgments were separated from the non-exclusive contribution labels, and 
and every  answer had to be accompanied by a direct quotation (`evidence') from the article.

\bigskip
\noindent 
{\bf Manual validation.}
We also conducted a manual validation on a subset of submissions. Because it is
infeasible to manually review all 32,944 submissions, we first performed a
keyword search over the parsed full texts using a comprehensive list of keywords
covering nearly all major AI model names. This screening step identified
approximately 5,000 candidate articles. We then sampled articles from this set
and manually completed the questionnaire for each sampled article. The manual
validation is still ongoing, and we have assembled a team of more than 10
volunteers to carry out this process.

\section{Results}

\subsection{A sharp recent increase, from a small base}\label{sec:trend}
We begin by examining how disclosed AI use changed over the study period.
The overall pattern is striking: AI use in mathematical research increased
rapidly between March and August 2026. Among all Mathematics submissions, the
share with confirmed AI use rose monotonically from 4.75\% in March to
24.14\% through August 20. The increase was even more pronounced for
substantive AI use, which rose from 1.39\% to 14.09\% over the same period.
Thus, by the end of the study period, both measures were several times higher
than at the beginning, indicating a rapid acceleration in disclosed AI
adoption in mathematical research.

\begin{figure}[t!]
\centering
\begin{minipage}{0.98\textwidth}
\centering

% ---------- Table 1 ----------
\captionof{table}{AI-use categories. ``All subcategories'' is the
de-duplicated number of articles carrying at least one label in the major
category; individual subcategory counts are non-exclusive.}
\label{tab:ai-categories-combined}

\setlength{\tabcolsep}{6.5pt}
\scalebox{.85}{
\begin{tabular}{@{}llrrrrrrr@{}}
\begin{tabular}{@{}llrrrrrrr@{}}
\toprule
Category & Subcategory & All & Mar & Apr & May & Jun & Jul & Aug$^*$ \\
\midrule
 & All articles & 32,944 & 5,265 & 5,097 & 5,525 & 6,106 & 6,464 & 4,487 \\
 & AI usage & 3,575 & 250 & 281 & 430 & 570 & 961 & 1,083 \\
\midrule
\multirow{5}{*}{Substantive} & All subcategories & 1,712 & 73 & 110 & 174 & 253 & 470 & 632 \\
 & Proof construction & 1,225 & 50 & 74 & 136 & 183 & 337 & 445 \\
 & Problem formulation & 146 & 7 & 9 & 15 & 29 & 41 & 45 \\
 & Formalization and verification & 427 & 23 & 31 & 34 & 67 & 108 & 164 \\
 & Other research assistance & 289 & 15 & 18 & 21 & 41 & 76 & 118 \\
\midrule
\multirow{5}{*}{Nonsubstantive} & All subcategories & 2,809 & 219 & 229 & 332 & 461 & 758 & 810 \\
 & Language and formatting & 2,108 & 165 & 169 & 227 & 321 & 585 & 641 \\
 & Code and computation & 1,229 & 97 & 111 & 158 & 219 & 308 & 336 \\
 & Literature and references & 524 & 33 & 26 & 47 & 66 & 157 & 195 \\
 & Other non-substantive assistance & 259 & 25 & 20 & 32 & 41 & 64 & 77 \\
\bottomrule
\end{tabular}
\end{tabular}
}

{\scriptsize $^*$ The August data do not cover the full month.}

\vspace{0.5cm}

% ---------- Figure 2 ----------
\begin{minipage}[t]{0.49\textwidth}
\centering
\includegraphics[width=\linewidth]{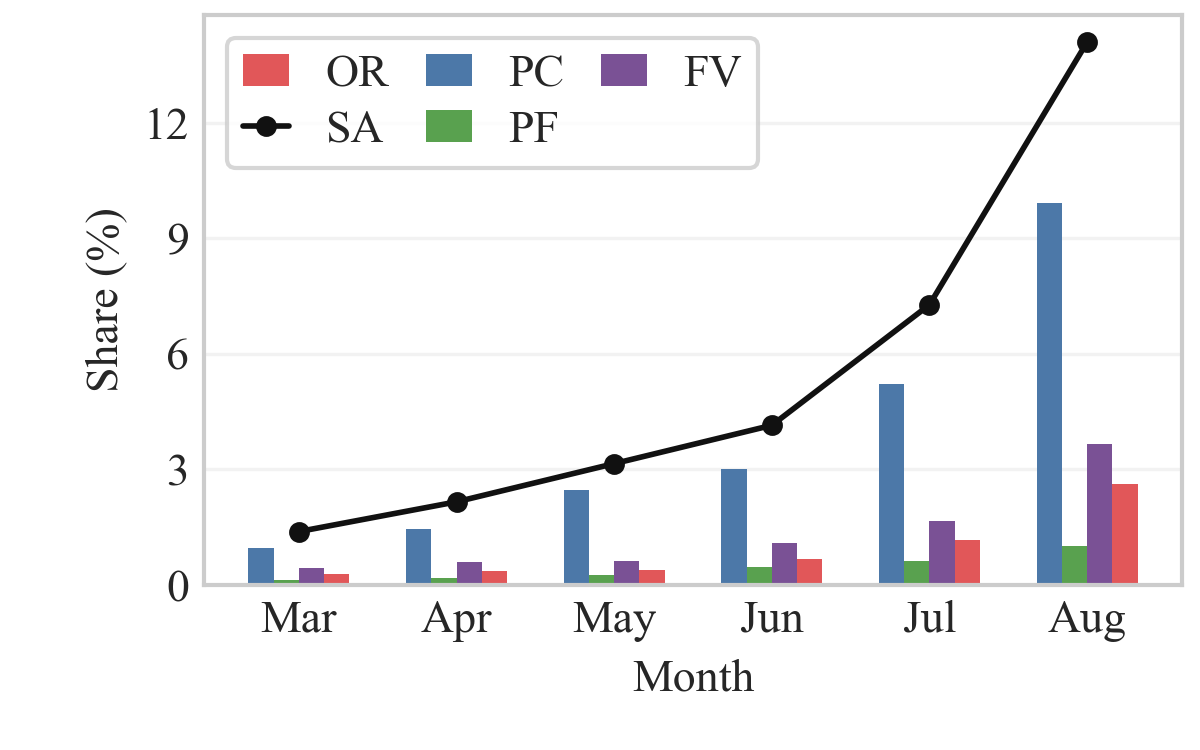}
\textbf{(a)} Substantive use
\end{minipage}
\hfill
\begin{minipage}[t]{0.49\textwidth}
\centering
\includegraphics[width=\linewidth]{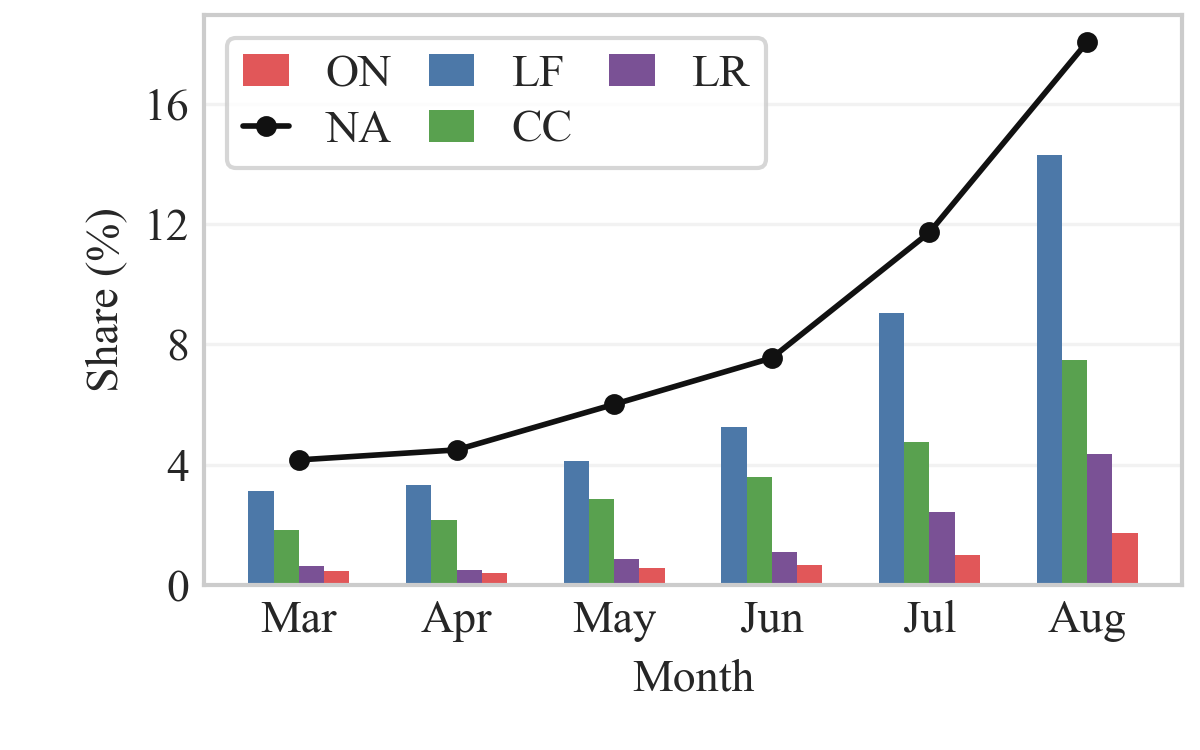}
\textbf{(b)} Non-substantive use
\end{minipage}

\vspace{.2cm}

\caption{Monthly AI-use category proportions. Panel (a) shows substantive
categories: PC (Proof construction), PF (Problem formulation), FV
(Formalization and verification), OR (Other research assistance), and SA
(substantive use in total). Panel (b) shows non-substantive categories:
LF (Language and formatting), CC (Code and computation), LR (Literature and
references), ON (Other non-substantive assistance), and NA (non-substantive
use in total).}
\label{fig:ai-usage-monthly}

\end{minipage}
\end{figure}

The nature of AI use is also heterogeneous. As shown in
Table~\ref{tab:ai-categories-combined}, language and formatting is by far the
most common individual use, indicating that much of the disclosed AI assistance
remains editorial rather than mathematical. At the same time, substantive uses
are already substantial. Proof construction is the largest substantive
subcategory, with 1,225 papers, followed by formalization and verification
(427), other research assistance (289), and problem formulation (146).
Because these subcategories are non-exclusive, a single paper may report
multiple forms of AI assistance.

Figure~\ref{fig:ai-usage-monthly} shows that the increase in AI use is broad
rather than driven by a single type of assistance. Both substantive and
non-substantive categories become more prevalent over the study period, and
most individual subcategories show substantial growth. In particular, the
rapid increase in proof construction and other substantive uses indicates that
the recent expansion of AI adoption extends beyond writing, coding, and other
non-substantive assistance to more direct participation in mathematical
research.

The rapid increase we observe is consistent with a broader rise in AI use
across scientific research. Large-scale studies have documented growing
AI-assisted writing and research across multiple disciplines \citep{liang2025quantifying,cheng2024ai,hao2026ai,he2026policies}.
These studies primarily capture linguistic signals or broad measures of AI
involvement. In contrast, our analysis uses explicit author disclosures and
distinguishes substantive mathematical contributions from non-substantive
assistance, suggesting that the recent growth of AI use in mathematics extends
beyond scientific writing to direct participation in mathematical research.

%The rapid increase we observe is consistent with a broader rise in AI use
%across scientific research. Using large corpora of scientific papers,
%\citet{liang2025quantifying} documented a steady increase in estimated
%LLM-modified writing, including in Mathematics, where the estimated prevalence
%reached approximately 9\% by 2024. Similarly, \citet{cheng2024ai} found
%evidence of increasing AI-generated or AI-revised text across multiple
%preprint disciplines, including the mathematical sciences.
%Other large-scale studies have likewise documented rapidly growing AI use
%across scientific fields \citep{hao2026ai,he2026policies}.
%These studies, however, primarily measure AI involvement through linguistic
%signals or broad indicators of AI-related research. In contrast, our analysis
%uses explicit author disclosures and distinguishes substantive mathematical
%contributions from non-substantive assistance. Our results therefore suggest
%that the recent growth of AI use in mathematics extends beyond scientific
%writing to direct participation in mathematical research.

\subsection{Which areas of mathematics are most affected?}\label{sec:fields}
The adoption of AI is highly uneven across mathematical fields.
Table~\ref{tab:field-primary} reports both the number of substantive-use papers
and the total number of submissions in each arXiv Mathematics subfield.
Combinatorics stands out with 397 substantive-use papers, followed by Number
Theory (135), Probability (110), and Algebraic Geometry (95). At the other
extreme, K-Theory and Homology, Quantum Algebra, and Symplectic Geometry each
have only four substantive-use papers. These differences already suggest
substantial heterogeneity in AI adoption across mathematical areas.

\begin{table}[tb!]
\centering
\caption{Primary-field results for all 31 arXiv Mathematics primary sub-fields. ``All'' gives the total number of submissions in each field, while ``AI-use'' gives the number in the confirmed-use sample. Substantive subcategory counts are non-exclusive.}
\label{tab:field-primary}

\setlength{\tabcolsep}{7pt}

\scalebox{.8}{
\begin{tabular}{p{4cm}p{2cm}|rrrr|rrrr}
\toprule
Sub-field & arXiv ID & All & AI-use & Substantive & Substantive \% & Proof & Form & Verify & Other \\
\midrule
Combinatorics & math.CO & 3,382 & 687 & 397 & 11.74\% & 305 & 35 & 99 & 49 \\
Number Theory & math.NT & 1,698 & 216 & 135 & 7.95\% & 102 & 12 & 40 & 18 \\
Optimization and Control & math.OC & 2,554 & 224 & 73 & 2.86\% & 52 & 2 & 18 & 12 \\
Probability & math.PR & 1,712 & 189 & 110 & 6.43\% & 86 & 6 & 21 & 13 \\
Numerical Analysis & math.NA & 2,195 & 182 & 42 & 1.91\% & 26 & 1 & 6 & 9 \\
Analysis of PDEs & math.AP & 3,035 & 167 & 67 & 2.21\% & 47 & 5 & 17 & 11 \\
Algebraic Geometry & math.AG & 1,462 & 144 & 95 & 6.50\% & 67 & 13 & 18 & 23 \\
Functional Analysis & math.FA & 988 & 104 & 52 & 5.26\% & 40 & 3 & 16 & 6 \\
Statistics Theory & math.ST & 785 & 92 & 42 & 5.35\% & 33 & 1 & 7 & 5 \\
Differential Geometry & math.DG & 1,254 & 90 & 49 & 3.91\% & 33 & 7 & 11 & 7 \\
Group Theory & math.GR & 657 & 78 & 43 & 6.54\% & 35 & 3 & 8 & 6 \\
Metric Geometry & math.MG & 305 & 72 & 40 & 13.11\% & 27 & 0 & 10 & 9 \\
Classical Analysis & math.CA & 547 & 73 & 43 & 7.86\% & 30 & 3 & 6 & 11 \\
Dynamical Systems & math.DS & 1,166 & 69 & 22 & 1.89\% & 14 & 2 & 10 & 4 \\
Mathematical Physics & math-ph & 817 & 48 & 17 & 2.08\% & 12 & 0 & 3 & 4 \\
Representation Theory & math.RT & 628 & 61 & 35 & 5.57\% & 25 & 6 & 8 & 5 \\
Geometric Topology & math.GT & 569 & 52 & 31 & 5.45\% & 19 & 5 & 7 & 7 \\
Logic & math.LO & 498 & 55 & 34 & 6.83\% & 24 & 2 & 11 & 4 \\
Operator Algebras & math.OA & 256 & 39 & 16 & 6.25\% & 9 & 3 & 2 & 7 \\
Algebraic Combinatorics & math.AC & 359 & 38 & 25 & 6.96\% & 16 & 2 & 6 & 5 \\
General Mathematics & math.GM & 274 & 46 & 14 & 5.11\% & 9 & 3 & 3 & 5 \\
Rings and Algebras & math.RA & 442 & 32 & 18 & 4.07\% & 11 & 1 & 5 & 5 \\
Complex Variables & math.CV & 455 & 29 & 16 & 3.52\% & 8 & 4 & 7 & 1 \\
Algebraic Topology & math.AT & 366 & 28 & 13 & 3.55\% & 6 & 0 & 3 & 6 \\
Category Theory & math.CT & 216 & 21 & 8 & 3.70\% & 6 & 3 & 1 & 1 \\
Spectral Theory & math.SP & 197 & 20 & 13 & 6.60\% & 10 & 2 & 1 & 3 \\
History and Overview & math.HO & 127 & 19 & 6 & 4.72\% & 2 & 1 & 3 & 1 \\
Quantum Algebra & math.QA & 216 & 14 & 4 & 1.85\% & 3 & 0 & 1 & 0 \\
Symplectic Geometry & math.SG & 167 & 11 & 4 & 2.40\% & 4 & 0 & 0 & 0 \\
General Topology & math.GN & 137 & 8 & 5 & 3.65\% & 5 & 0 & 0 & 0 \\
K-Theory and Homology & math.KT & 52 & 4 & 4 & 7.69\% & 4 & 0 & 0 & 0 \\
\bottomrule
\end{tabular}
}
\setlength{\tabcolsep}{6pt}
\end{table}

The picture changes somewhat after accounting for field size. Metric Geometry
has the highest substantive-use rate, with 40 of 305 papers (13.11\%),
followed by Combinatorics at 11.74\%. Number Theory and Classical Analysis also
have relatively high rates, at 7.95\% and 7.86\%, respectively. In contrast,
some large fields, such as Numerical Analysis and Analysis of PDEs, contain
many AI-use papers in absolute terms but have much lower substantive-use rates,
at 1.91\% and 2.21\%.

These differences suggest that current AI adoption is not simply proportional
to the volume of research produced in a field. Some areas appear more receptive
to substantive AI assistance than others. This is consistent with recent observations
that automated reasoning has been particularly successful in areas such as
combinatorics, algebra, number theory, and discrete geometry
\citep{avigad2026mathematicians}. More broadly, problem solving, proof
generation, and formal verification have been identified as aspects of
mathematical practice especially susceptible to AI progress
\citep{tao2026mathematics,commelin2026shaping}. 
Differences in community norms,
disclosure practices, and the composition of early adopters may also contribute.
Our observational data, however, do not allow us to distinguish among these
explanations.

%\begin{center}
%\small
%\begin{longtable}{>{\RaggedRight\arraybackslash}p{3.5cm} >{\RaggedRight\arraybackslash}p{2.3cm} >{\RaggedRight\arraybackslash}p{1.6cm}}
%\caption{Open-problem field distribution by primary field.}\label{tab:open-primary}\\
%\toprule
%Field & arXiv & Problem records \\
%\midrule
%\endfirsthead
%\toprule
%Field & arXiv & Problem records \\
%\midrule
%\endhead
%Combinatorics & math.CO & 772 \\
%Number Theory & math.NT & 240 \\
%Probability & math.PR & 169 \\
%Algebraic Geometry & math.AG & 122 \\
%Geometric Topology & math.GT & 114 \\
%Optimization and Control & math.OC & 109 \\
%Analysis of PDEs & math.AP & 96 \\
%Differential Geometry & math.DG & 88 \\
%Functional Analysis & math.FA & 78 \\
%Group Theory & math.GR & 72 \\
%Metric Geometry & math.MG & 72 \\
%Classical Analysis & math.CA & 65 \\
%Dynamical Systems & math.DS & 65 \\
%Representation Theory & math.RT & 62 \\
%Statistics Theory & math.ST & 61 \\
%Logic & math.LO & 51 \\
%Numerical Analysis & math.NA & 47 \\
%Algebraic Combinatorics & math.AC & 44 \\
%General Mathematics & math.GM & 35 \\
%Complex Variables & math.CV & 29 \\
%Mathematical Physics & math-ph & 28 \\
%Spectral Theory & math.SP & 28 \\
%Operator Algebras & math.OA & 24 \\
%Algebraic Topology & math.AT & 22 \\
%Rings and Algebras & math.RA & 22 \\
%History and Overview & math.HO & 15 \\
%Category Theory & math.CT & 9 \\
%General Topology & math.GN & 9 \\
%Quantum Algebra & math.QA & 7 \\
%Symplectic Geometry & math.SG & 6 \\
%K-Theory and Homology & math.KT & 1 \\
%\bottomrule
%\end{longtable}
%\end{center}

\subsection{What open problems have been solved?}

\begin{figure}[tb]
\centering
\includegraphics[width=\textwidth, trim=20 60 20 0, clip=true]{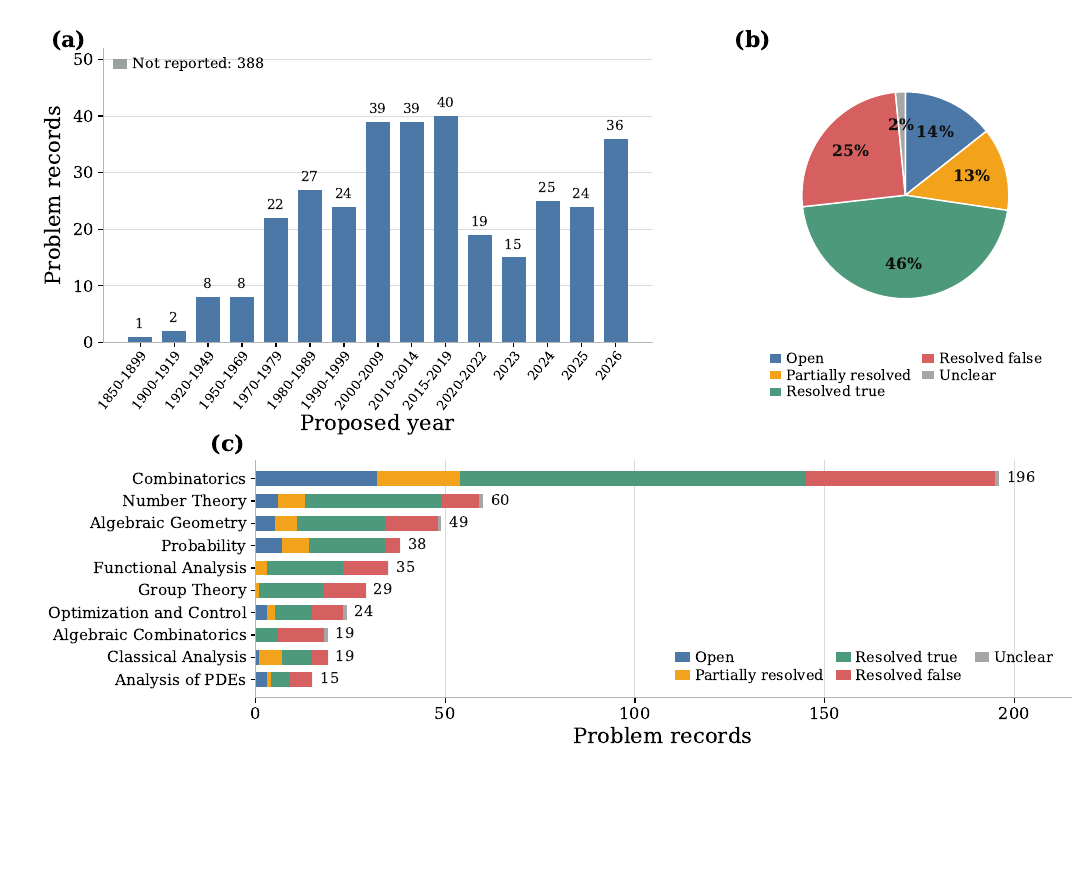}
\caption{Named open-problem analysis. Panel (a) shows the distribution of problem records by audited proposed-year interval among records with a reported proposed year; the 388 records for which the proposed year was not reported are omitted from the bars and noted in the legend. Panel (b) shows the overall status distribution. Panel (c) shows the ten primary fields with the largest numbers of open-problem records, with bars segmented by status classification.}
\label{fig:open-problem-analysis}
\end{figure}

A particularly interesting question is whether AI is being used only for routine mathematical
tasks or is already contributing to work on open research problems. Among the
substantive-use papers, our review identifies 717 named open-problem records.
These records span a wide range of mathematical fields and include problems
proposed across many decades, as summarized in
Figure~\ref{fig:open-problem-analysis}.

Of the 717 open-problem records, 103 remain {\it open} and 93 are {\it partially
resolved}. A much larger group, 510 records, are labeled as fully resolved:
329 as {\it resolved true} and 181 as {\it resolved false}; the remaining 11
are classified as {\it unclear}. These labels are assigned by the LLM-based
review from the authors' own descriptions in the manuscripts and do not
constitute independent mathematical verification of the claimed resolutions.
Among the records labeled as fully resolved, proofs of the conjectured
statement are more common than counterexamples or disproofs. These results
indicate that substantive AI use is already extending beyond established
mathematics to work on unresolved research questions.

The open-problem records are also highly concentrated across fields.
Combinatorics accounts for by far the largest number, followed by Number
Theory, Algebraic Geometry, and Probability; see
Figure~\ref{fig:open-problem-analysis}(c). This pattern broadly mirrors the
field distribution of substantive AI use reported above, suggesting that the
areas currently adopting AI most actively are also among those where AI is
being applied most frequently to open research problems.

The proposed years also span a remarkably wide range; see
Figure~\ref{fig:open-problem-analysis}(a). Among the 329 records with a
reported proposed year, 92 date to before 2000, including 68 before 1990,
while many others concern much more recent problems. In particular, 36 records
correspond to problems proposed in 2026 itself. Thus, the sample includes both
long-standing conjectures and newly posed research questions. Because the
proposed year is not reported for 388 records, however, this distribution
should be interpreted only among problems with an identifiable proposed date.

\begin{center}
\scriptsize
\begin{longtable}{>{\raggedright\arraybackslash}p{0.3cm} >{\raggedright\arraybackslash}p{1.5cm} >{\raggedright\arraybackslash}p{4.1cm} >{\raggedright\arraybackslash}p{2.3cm} >{\centering\arraybackslash}p{1cm} >{\centering\arraybackslash}p{1.5cm} >{\raggedright\arraybackslash}p{4.55cm}}
\caption{The 50 earliest proposed open-problem records satisfying three
criteria: substantive AI use, an available proposed year, and a reported full
resolution. Records are ordered by proposed year.}\label{tab:conjectures-top50}\\
\toprule
 & arXiv ID & Conjecture & Sub-fields & Proposed & Outcome & AI model \\
\midrule
\endfirsthead
\toprule
Rank & arXiv ID & Conjecture & Sub-fields & Proposed & Outcome & AI model \\
\midrule
\endhead
1 & 2604.09808 & Ramanujan Nagell theorem & math.NT & 1913 & proved & Anthropic Claude Code; Aristotle AI Aristotle \\
2 & 2608.15904 & Characterization of entrywise positivity preservers in fixed dimension & math.CA; math.FA & 1925 & proved & OpenAI ChatGPT 5.6 Sol \\
3 & 2608.13536 & Banach isometric conjecture & math.FA; math.DG; math.MG & 1932 & proved & OpenAI ChatGPT 5.5 Pro; OpenAI ChatGPT 5.6 Pro; OpenAI GPT 5.6 Sol \\
4 & 2608.12561 & Borsuk's problem & math.MG; math.CO & 1933 & disproved & OpenAI GPT-5.6 Sol \\
5 & 2608.15681 & Erdos Problem & math.NT & 1943 & proved & OpenAI ChatGPT-5.6 \\
6 & 2605.20695 & Erdos unit distance conjecture & math.CO; math.NT & 1946 & disproved & OpenAI Chat GPT \\
7 & 2605.28793 & Erdos conjecture that r(s,k) is at least k\textasciicircum{}(s-1) up to polylogarithmic factors & math.CO & 1947 & proved & not specified \\
8 & 2607.24483 & Bellman’s lost-in-a-forest problem & math.MG & 1956 & proved & Anthropic Claude Fable 5; OpenAI GPT 5.6 Sol; Anthropic Claude Opus 5 \\
9 & 2607.14071 & Erdos Conjecture On Complete Sets & math.NT; math.CO & 1961 & proved & OpenAI ChatGPT 5.6 \\
10 & 2603.29961 & Whether the fractional parts of alpha times the primes are well-distributed & math.CO; math.NT & 1964 & proved & Openai GPT 5.4 Pro \\
11 & 2607.12935 & Lindenstrauss retraction problem & math.FA & 1964 & disproved & OpenAI ChatGPT 5.6 Pro \\
12 & 2605.00301 & Erdos Sarkozy Szemeredi Problem 1196 & math.NT; math.CO; math.PR & 1966 & proved & OpenAI GPT-5.4 Pro; OpenAI GPT-5.5 Pro; OpenAI Codex; Math Inc Gauss; OpenAI ChatGPT \\
13 & 2605.00301 & Erdos Sarkozy Szemeredi Problem 1217 & math.NT; math.CO; math.PR & 1966 & proved & OpenAI GPT-5.4 Pro; OpenAI GPT-5.5 Pro; OpenAI Codex; Math Inc Gauss; OpenAI ChatGPT \\
14 & 2608.02852 & Forsythe conjecture for restart length two & math.NA; math.DS; math.OC & 1968 & proved & GPT-5.6 Sol Ultra \\
15 & 2607.22515 & Rota's conjecture for flats & math.CO & 1970 & disproved & OpenAI ChatGPT 5.6 Pro \\
16 & 2603.19215 & R-equivalence of Manin's diagonal cubic surface over Q 2 zeta 3 & math.AG; math.NT & 1972 & proved & Google DeepMind AlphaEvolve; Google DeepMind Gemini 3 Pro; Google DeepMind Gemini 3 Deep Think; Google DeepMind Gemini 2.5 Pro; Google DeepMind Gemini 3.1 Pro \\
17 & 2607.14140 & Cycle Double Cover Conjecture & math.CO & 1973 & proved & OpenAI Codex \\
18 & 2605.23837 & Gale first move uniqueness question & math.CO & 1974 & proved & not specified \\
19 & 2606.24872 & Erdos Problem 768 & math.NT & 1974 & proved & OpenAI ChatGPT; Harmonic Aristotle \\
20 & 2608.07388 & Sigma Lebesgue completion problem & math.FA & 1974 & disproved & OpenAI GPT-5.5; Anthropic Claude Opus 4.6 \\
21 & 2605.00301 & Erdos Primitive Set Conjecture Problem 164 & math.NT; math.CO; math.PR & 1976 & proved & OpenAI GPT-5.4 Pro; OpenAI GPT-5.5 Pro; OpenAI Codex; Math Inc Gauss; OpenAI ChatGPT \\
22 & 2605.28781 & sum-product conjecture & math.NT; math.CO & 1976 & disproved & OpenAI GPT-5.5 Pro \\
23 & 2607.20525 & The Erdos Szemeredi Sum Product Conjecture over R & math.CO; math.NT & 1976 & disproved & OpenAI GPT-5.5 Pro; OpenAI GPT-5.5; OpenAI GPT-5.5 Codex \\
24 & 2606.13925 & Grothendieck vanishing theorem & math.AG & 1977 & proved & Anthropic Claude Code; Anthropic Claude; OpenAI Codex; Aristotle \\
25 & 2608.16977 & Erdos and Straus question on divisibility among binomial coefficients & math.CO & 1977 & proved & gpt-oss-120b; gemini-3.5-flash; gemini-3.1-pro; gpt-5.5; opencode agent \\
26 & 2604.11220 & weaker form of Yau s uniformisation conjecture & math.DG & 1978 & proved & OpenAI ChatGPT pro 5.2 \\
27 & 2608.08799 & Delta Conjecture & math.CO & 1978 & proved & OpenAI ChatGPT \\
28 & 2605.08542 & Erdos Problem 690 & math.NT & 1979 & proved & OpenAI gpt-5.4-pro; OpenAI gpt-5.5; Anthropic Claude Opus 4.7 \\
29 & 2608.08118 & Bernhart Kainen dispersability conjecture & math.CO & 1979 & disproved & OpenAI GPT-5.5; Anthropic Claude Opus 4.7; Google Gemini 3 Flash; Google Gemini 3.1 Pro; Anthropic Claude Sonnet 4.6 \\
30 & 2607.13165 & Brualdi Problem 3.7 Hamiltonicity of interchange graphs & math.CO & 1980 & proved & Anthropic Claude; OpenAI GPT/Codex \\
31 & 2607.15419 & Erdos and Graham question on large sets with a+b not dividing 2ab & math.NT; math.CO & 1980 & disproved & ChatGPT \\
32 & 2608.02126 & The Kalton--Peck hyperplane problem & math.FA & 1980 & proved & OpenAI GPT-5.5 \\
33 & 2608.08953 & Planar Berenstein Conjecture & math.AP; math-ph; math.NA; math.SP & 1980 & disproved & OpenAI ChatGPT 5.6 \\
34 & 2605.03066 & Odifreddi Problem 3 & math.LO & 1981 & disproved & Google DeepMind Gemini Deep Think \\
35 & 2606.24878 & Whether g r is less than 3 r & math.CO & 1981 & proved & OpenAI ChatGPT 5.5 Pro; Harmonic’s Aristotle \\
36 & 2608.08933 & Kim Roush Conjecture & math.CO & 1981 & proved & OpenAI GPT-5.6-sol; Anthropic Claude Fable 5 \\
37 & 2608.16040 & Chudnovsky Conjecture & math.AG; math.AC & 1981 & proved & OpenAI ChatGPT-5.6 Sol \\
38 & 2607.20816 & Erdos Problem 906 & math.CV & 1982 & proved & OpenAI GPT-5.6 Sol \\
39 & 2608.02327 & Connes’ rigidity conjecture for ICC property (T) groups. & math.OA; math.DS; math.GR & 1982 & disproved & GPT-5.6 Sol; Codex; Danus multi-agent research system \\
40 & 2608.16040 & Demailly Conjecture & math.AG; math.AC & 1982 & proved & OpenAI ChatGPT-5.6 Sol \\
41 & 2608.14438 & Conjecture 1.2 Volume Growth & math.DG & 1986 & proved & OpenAI ChatGPT 5.6 Sol Ultra; OpenAI Codex \\
42 & 2608.14507 & Gromov's conjectured codimension-two volume growth estimate & math.DG & 1986 & proved & OpenAI GPT-5.6 Sol with Ultra reasoning efort; OpenAI Codex \\
43 & 2608.18068 & Stein's dimension free weak type (1,1) problem & math.CA; math.AP & 1986 & proved & OpenAI ChatGPT GPT 5.6 Sol; OpenAI Codex CLI; Anthropic Claude Opus 5.0; OpenAI Danus; OpenAI Rethlas \\
44 & 2607.00359 & Katz Problem On Equivalence Of Generalized Airy Operators & math.RA & 1987 & proved & MechMath Agent Team (MMAT) \\
45 & 2608.13025 & Foregger Sinkhorn tie point conjecture & math.CO & 1987 & disproved & OpenAI GPT-5.6-sol; Anthropic Claude Fable 5 \\
46 & 2607.22988 & Stanley's rankwise lower-bound conjecture & math.CO & 1988 & disproved & TARS agent system \\
47 & 2607.24541 & Stanley Problem 4 on Differential Posets & math.CO & 1988 & disproved & TARS agent system \\
48 & 2606.23659 & Erdos Problem & math.CO & 1989 & proved & OpenAI ChatGPT; OpenAI Codex; Harmonic Aristotle \\
49 & 2607.19192 & Donoho Stark conjecture & math.CA; math.FA & 1989 & proved & OpenAI GPT 5.6 Sol; OpenAI GPT-5.5 Pro \\
50 & 2608.15515 & Talagrand's convolution conjecture & math.PR & 1989 & proved & Odin Automatic AI Research Agent \\
\bottomrule
\end{longtable}
\end{center}

Table~\ref{tab:conjectures-top50} provides a closer look at selected
open-problem records. We restrict attention to records that satisfy three
criteria: the corresponding paper involves substantive AI use, the proposed
year of the problem is available, and the problem is classified as fully
resolved. Among these records, we report the first 50 when ordered by proposed
year, from earliest to most recent. The ``arXiv ID'' column identifies the
corresponding submission; 
%``Conjecture'' gives the name or description of the problem; 
and ``Sub-fields'' lists all Mathematics subcategories associated with the
submission, excluding non-mathematics categories. 
%``Proposed'' gives the reported year in which the problem was posed; ``Outcome'' indicates whether the reported resolution proves or disproves the conjectured statement; and ``AI model'' lists the AI systems disclosed in the manuscript.

Table~\ref{tab:conjectures-top50} also has substantial overlap with existing
online collections of open mathematical problems. Most notably, 14 of the 50
entries are explicitly Erd\H{o}s-related. These include Erd\H{o}s Problem
\#90, the unit-distance problem; Problem \#986, concerning lower bounds for
Ramsey numbers; Problems \#1196 and \#1217 of Erd\H{o}s--S\'ark\"ozy--Szemer\'edi;
and Problems \#768, \#164, \#690, and \#906. The table also contains the
Erd\H{o}s conjecture on complete sets, the Erd\H{o}s--Szemer\'edi sum-product
conjecture, an Erd\H{o}s--Straus question on binomial coefficients, and an
Erd\H{o}s--Graham divisibility problem. Thus, the Erd\H{o}s Problems
collection \citep{bloomErdosProblems} provides a particularly visible point of
overlap between our observational sample and existing repositories of
research-level mathematical problems.

%It is useful to compare Table~\ref{tab:conjectures-top50} with several recent
%collections of research-level mathematical problems used to evaluate AI.
%The Erd\H{o}s Problems database provides a large historical collection,
%primarily in combinatorics and number theory \citep{bloomErdosProblems},
%and has recently served as a testbed for systematic AI-assisted problem
%solving \citep{feng2026erdos}.
%FrontierMath: Open Problems instead curates a smaller
%set of particularly notable unsolved problems with computational verifiers,
%while Formal Conjectures collects thousands of research-level statements
%formalized in Lean, including more than one thousand open conjectures
%\citep{frontiermath2026open,firsching2026formal}. The \emph{First Proof}
%project takes yet another approach, using previously unpublished research
%problems and independent expert evaluation \citep{abouzaid2026firstproof}.
%In contrast, Table~\ref{tab:conjectures-top50} is not a benchmark: it is an
%observational sample drawn from naturally occurring arXiv manuscripts in which
%authors disclose substantive AI use. It therefore complements these curated
%problem sets by showing where AI-assisted work on open problems is already
%appearing in actual mathematical research.

These findings should, however, be interpreted cautiously. The resolution
status is inferred by our LLM-based review from the authors' descriptions in
the manuscripts and does not constitute independent mathematical verification.
A paper classified as resolving an open problem may later be found to contain
an error, overlap with prior work, or require further expert validation.
Accordingly, we interpret these results as evidence about the kinds of
open-problem claims appearing in AI-assisted mathematics, rather than as a
definitive count of previously open problems solved by AI.

\FloatBarrier
\subsection{Who is using AI for mathematical research?}\label{sec:authors}

We next investigate who is participating in substantive AI-assisted mathematical
research. The 1,712 substantive-use manuscripts in our sample involve 2,764
canonical authors and 744 university-level institutions. Institution names
were normalized to merge equivalent variants, such as ``University of
Cambridge'' and ``Cambridge University.'' We then examine how unevenly
substantive AI use is distributed across authors, institutions, and countries.

%\begin{table}[H]
%\centering
%\small
%\caption{Distribution of confirmed-use authors by number of papers per author. Paper counts with no authors are omitted; the entries sum to 5,818 canonical authors.}
%\label{tab:authors-by-paper-count}
%\begin{tabular}{r r @{\hspace{1.15cm}} r r @{\hspace{1.15cm}} r r}
%\toprule
%Papers & Authors & Papers & Authors & Papers & Authors \\
%\midrule
%30 & 1 & 21 & 1 & 17 & 2 \\
%16 & 3 & 14 & 2 & 13 & 1 \\
%12 & 1 & 10 & 2 & 9 & 5 \\
%8 & 6 & 7 & 12 & 6 & 13 \\
%5 & 38 & 4 & 49 & 3 & 195 \\
%2 & 647 & 1 & 4,840 & & \\
%\bottomrule
%\end{tabular}
%\end{table}

\begin{figure}[tb!]
\centering
\includegraphics[width=.5\textwidth, trim=0 0 0 20, clip=true]{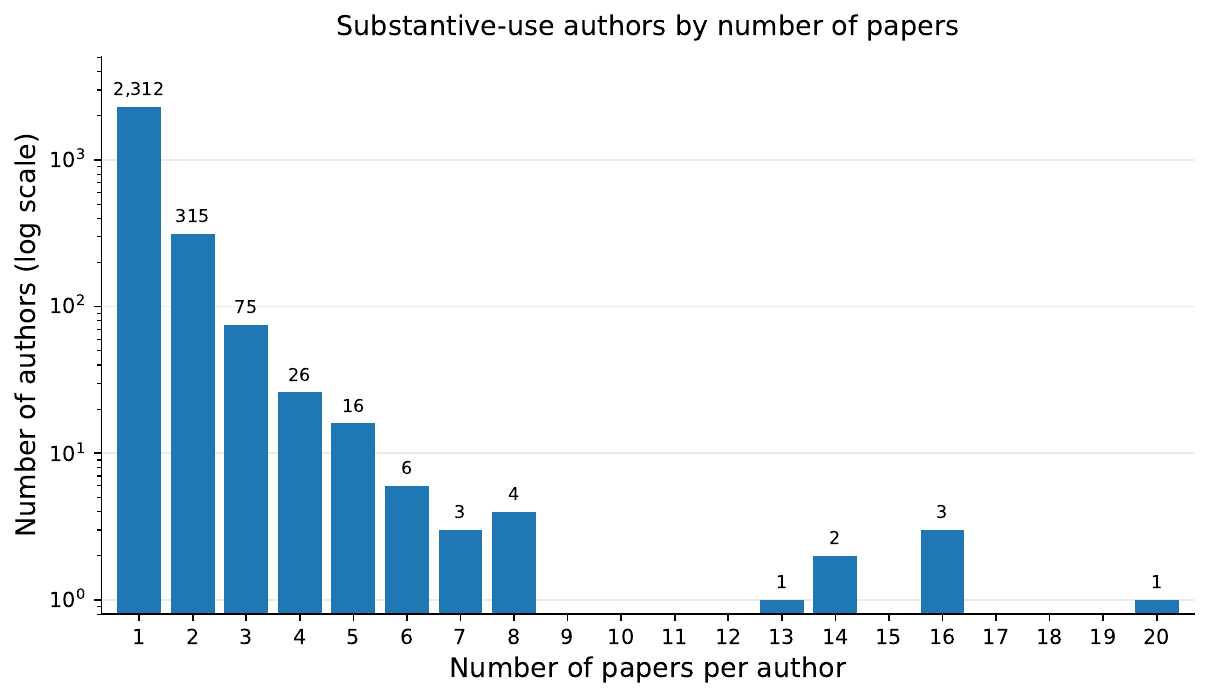}
\caption{Histogram of substantive-use manuscripts per canonical author. The vertical axis is on a logarithmic scale.}
\label{fig:substantive-author-histogram}
\end{figure}

\begin{figure}[tb!]
\centering
\includegraphics[width=0.95\textwidth]{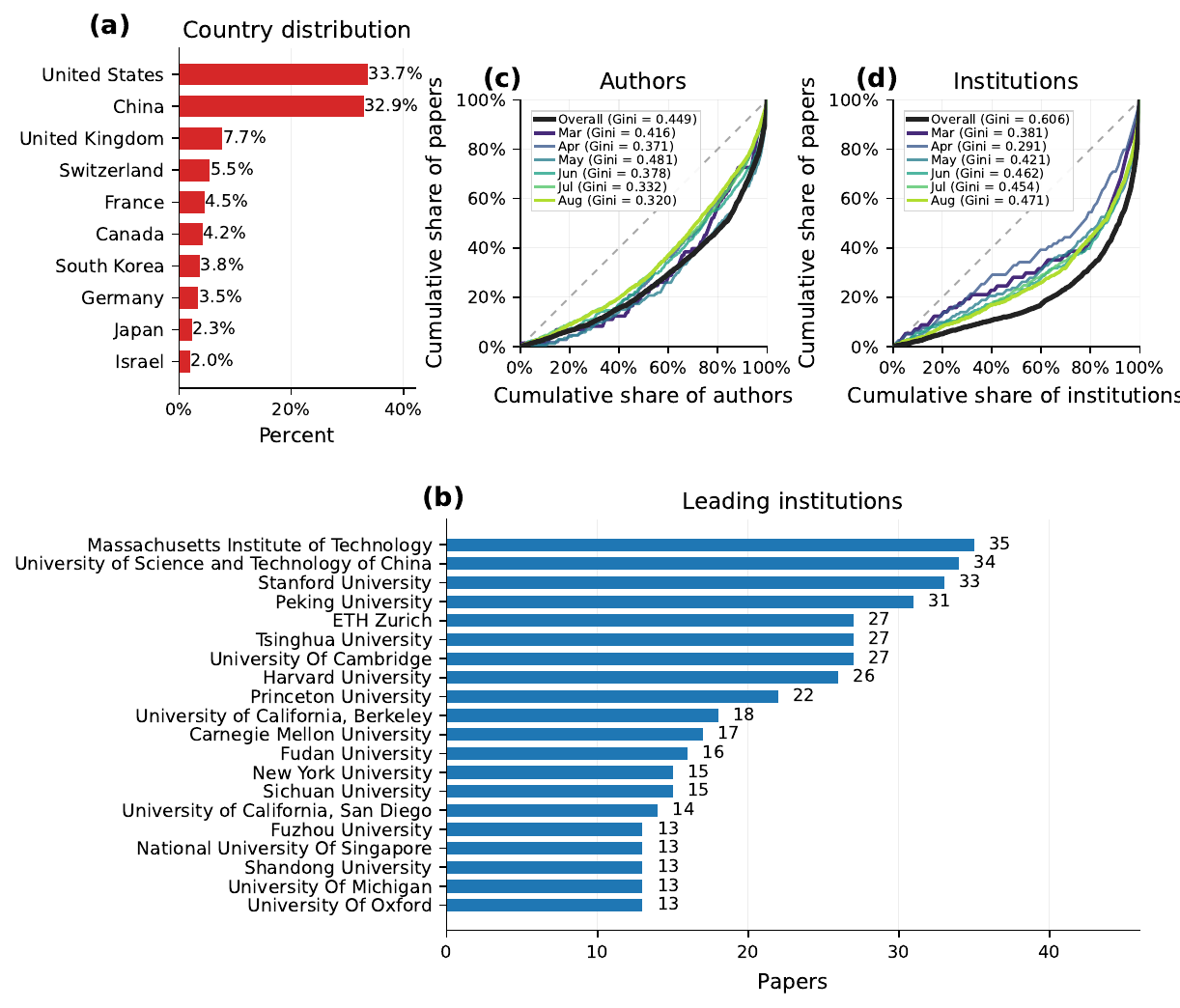}
\caption{Country, institution, and concentration summaries for the substantive-use sample. Panel (a) shows the ten largest recognized country shares among author records; multi-country affiliations are split fractionally, unmatched author records are excluded, and percentages are normalized over recognized country weight. Panel (b) reports the leading normalized school-level institutions by unique substantive-use manuscripts, with each manuscript counted at most once per institution. Panels (c) and (d) show author and institution concentration in conventional Lorenz orientation, overlaying the pooled sample with monthly curves from March through August 2026; each legend reports the corresponding Gini coefficient. For the institution curves, unknown or non-school labels and papers without a known school are excluded.}
\label{fig:author-overview}
\label{fig:author-institution-gini}
\end{figure}

The distribution is highly uneven across authors; see
Figure~\ref{fig:substantive-author-histogram}. Among the 2,764 canonical
authors, 83.6\% appear on only one substantive-use manuscript, and 11.4\%
appear on two. Thus, more than 95\% of authors appear on at most two
substantive-use papers. Only a small minority contribute repeatedly: 75
authors appear on three papers and 26 on four, while a handful of particularly
active authors appear on more than ten manuscripts, with the maximum reaching
20. Panel~(c) of Figure~\ref{fig:author-institution-gini} quantifies this
concentration using Lorenz curves and Gini coefficients. In the pooled
substantive-use sample, the author-level Gini coefficient is 0.449. The
month-specific coefficients are 0.416, 0.371, 0.481, 0.378, 0.332, and 0.320
from March through August, respectively. Thus, with the exception of May,
monthly concentration is lower than in the pooled sample. A natural reason is
that pooling across months accumulates repeated contributions by the same
authors: an author who appears only once or twice within any given month may
appear many times over the full study period. Such repeated participation
increases the dispersion of cumulative manuscript counts across authors and
therefore raises the pooled Gini coefficient.

The geographic distribution is also highly concentrated; see Figure~\ref{fig:author-overview}(a). To account for
differences in team size, we use weighted author counts: if a manuscript has
$m$ authors, each author contributes a weight of $1/m$. These author weights
are then assigned to countries based on the authors' recognized affiliations, where multi-country affiliations are split fractionally, while unmatched author records are excluded from both the numerator and denominator; the reported percentages are therefore normalized over recognized country assignments.   
%with multi-country affiliations split fractionally when necessary. 
Among the
resulting recognized country weights, the United States accounts for 33.7\%
and China for 32.9\%, together representing about two-thirds of the total.
The United Kingdom is a distant third at 7.7\%, followed by Switzerland
(5.5\%), France (4.5\%), and Canada (4.2\%). Each of the remaining countries
accounts for less than 4\%.

Institutional participation is even more concentrated; see
Figure~\ref{fig:author-overview}(b) and (d). Among the 744 
university-level institutions, the Massachusetts Institute of Technology
appears on the largest number of substantive-use manuscripts (35), followed
closely by the University of Science and Technology of China (34), Stanford
University (33), and Peking University (31). ETH Zurich, Tsinghua University,
and the University of Cambridge each appear on 27 manuscripts. Each manuscript
is counted at most once for a given institution, although a manuscript with
multiple institutional affiliations may contribute to multiple institutions.
Panel~(d) further shows that the pooled institution-level Gini coefficient is
0.606, substantially higher than the corresponding author-level coefficient
of 0.449. The monthly institution-level coefficients are 0.381, 0.291, 0.421,
0.462, 0.454, and 0.471 from March through August, respectively, all below the
pooled value. As with authors, pooling across months accumulates repeated
participation by the same institutions. Institutions that appear regularly
throughout the study period therefore separate increasingly from those that
appear only occasionally, producing greater concentration in the pooled
sample.

\begin{figure}[tb!]
\centering
\includegraphics[width=0.8\textwidth]{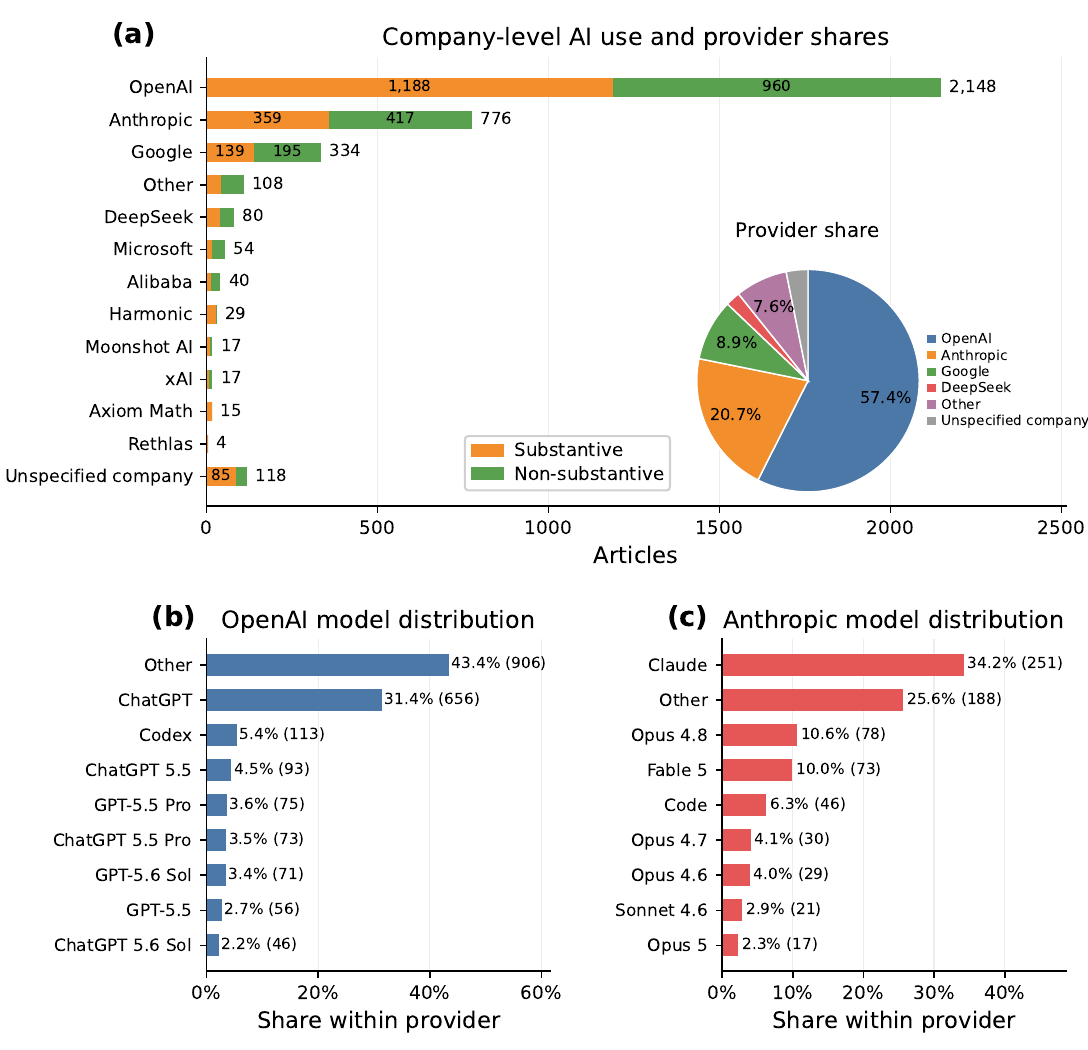}
\caption{Provider and model distributions in the confirmed-use sample. Panel (a) shows the number of manuscripts associated with each provider, partitioned into substantive-use and non-substantive-only manuscripts. The inset pie chart normalizes provider--manuscript associations across providers, so its percentages differ from the manuscript-level prevalence reported in the text. Panels (b) and (c) show the distributions of reported model labels within OpenAI and Anthropic, respectively. Provider and model mentions are non-exclusive across manuscripts because a manuscript may name multiple providers or models.}
\label{fig:company-overall-bar}
\end{figure}

\subsection{Which AI systems are being used?}\label{sec:models}
The use of AI systems is concentrated among a small number of
providers. As show in Panel~(a) of Figure~\ref{fig:company-overall-bar}, OpenAI models appear in 2,148 of the 3,575 confirmed-use
manuscripts (60.08\%), followed by Anthropic in 776 (21.71\%), Google in
334 (9.34\%), and DeepSeek in 80 (2.24\%). Other providers appear much less
frequently. Because a manuscript may report using models from more than one
provider, these counts and percentages are non-exclusive and need not sum to
the total number of confirmed-use manuscripts.

Panels~(b)--(c) of Figure~\ref{fig:company-overall-bar} show that reported
usage is also concentrated within providers. For both OpenAI and Anthropic, a
small number of product or model labels account for a large share of the
reported mentions. Some labels, such as ``ChatGPT'' and ``Claude,'' are broad
product names that may encompass multiple underlying model versions, whereas
others identify a specific model more precisely. The within-provider
distributions should therefore be interpreted as distributions of
author-reported system names rather than exact model-level market shares.

%The substantive-use pattern broadly follows the overall provider distribution, with OpenAI-linked manuscripts accounting for the largest substantive-use count, followed by Anthropic and Google. This does not imply that one provider's models are intrinsically more capable of mathematical research: observed use depends on model availability, pricing, familiarity, workflow integration, and disclosure practices. The results nevertheless show that the current wave of AI-assisted mathematics is being shaped disproportionately by a small number of model providers.

\subsection{Article metrics across the archive and confirmed-use groups}\label{sec:characteristics}

\begin{figure}[H]\centering
\begin{minipage}[t]{0.32\textwidth}\centering\includegraphics[width=.9\linewidth, trim= 0 20 0 0, clip=true]{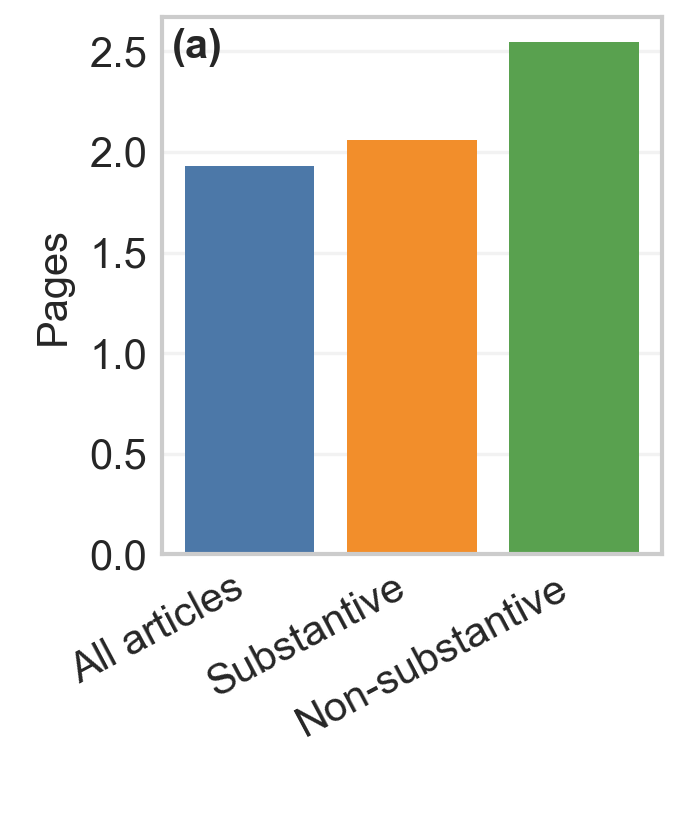}\end{minipage}\hfill
\begin{minipage}[t]{0.32\textwidth}\centering\includegraphics[width=.9\linewidth, trim= 0 20 0 0, clip=true]{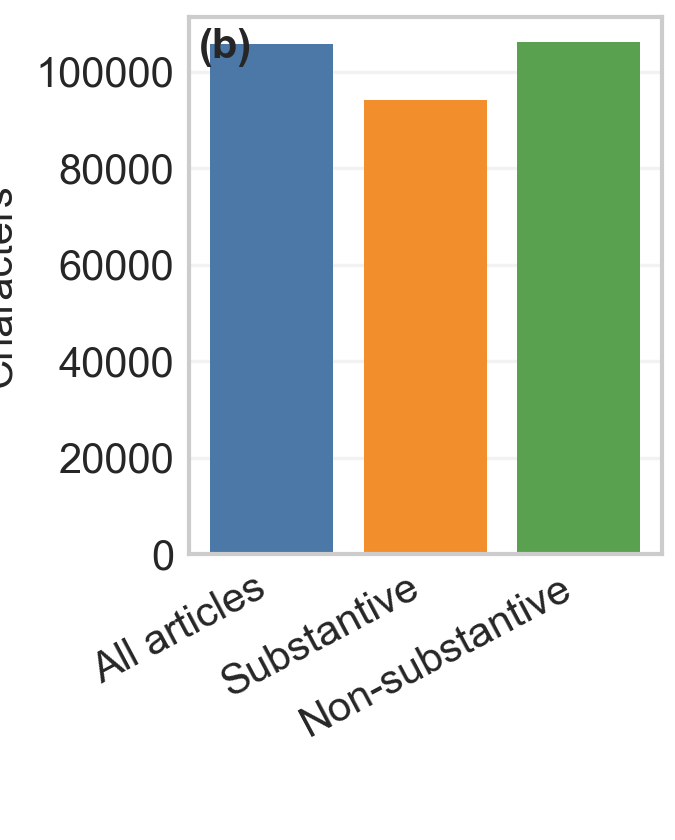}\end{minipage}\hfill
\begin{minipage}[t]{0.32\textwidth}\centering\includegraphics[width=.9\linewidth, trim= 0 20 0 0, clip=true]{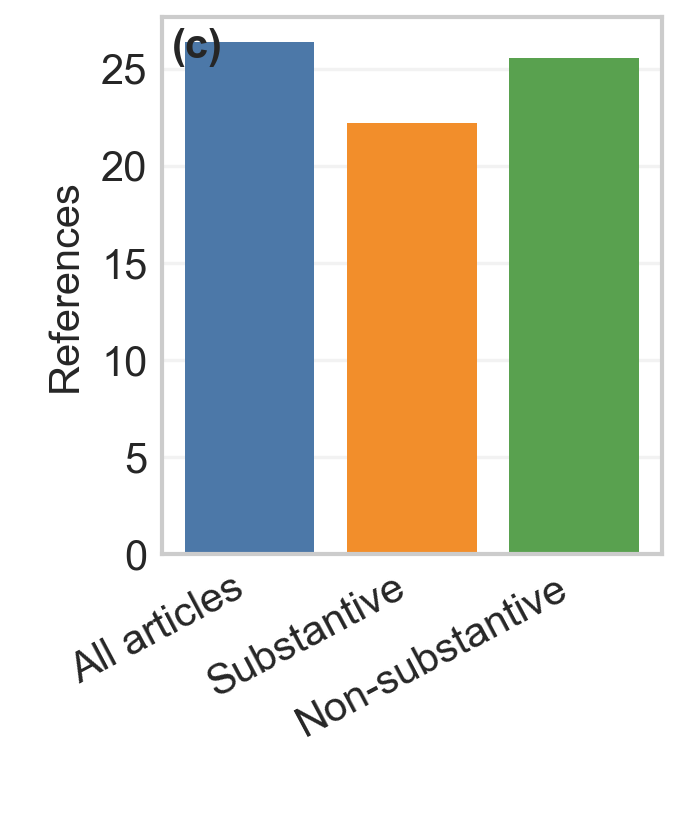}\end{minipage}
%\vspace{-.3cm}
\caption{Mean article metrics for all 32,944 Mathematics submissions (blue), the 1,712 confirmed-use manuscripts with substantive use (orange), and the remaining 1,863 confirmed-use manuscripts with no substantive-use label (green). The blue baseline is the full archive; the orange and green groups form a mutually exclusive partition of the confirmed-use sample.}\label{fig:article-metrics-latest}\end{figure}

Finally, Figure~\ref{fig:article-metrics-latest} compares several observable
article characteristics across the full archive and the two mutually exclusive
confirmed-use groups. The full archive contains all 32,944 Mathematics
submissions, while the confirmed-use sample is divided into 1,712
substantive-use papers and 1,863 non-substantive-only papers. The full archive
therefore serves as a descriptive baseline and overlaps with both confirmed-use
groups. 
On average, papers in the full archive contain 1.93 pages, 105,756.9 extracted characters, and 26.36 references. The corresponding means are 2.06 pages, 94,032.3 characters, and 22.22 references for substantive-use papers, compared with 2.54 pages, 106,050.6 characters, and 25.54 references for non-substantive-only papers.

Within the confirmed-use sample, substantive-use papers are therefore shorter
on average than non-substantive-only papers, by approximately 0.48 pages and
12,018 extracted characters, and contain about 3.3 fewer references. The
permutation comparisons indicate statistically detectable differences in page
count and extracted text length, whereas the difference in reference count is
not statistically significant. 
%Thus, the clearest descriptive difference between the two confirmed-use groups is in manuscript length rather than the number of references.

%Relative to the non-substantive-only group, substantive-use manuscripts are shorter on all three displayed averages: by about 0.49 pages, 12,018 extracted characters, and 3.31 references. These comparisons are descriptive and are not adjusted for differences in field, authorship, or other manuscript characteristics.

One possible interpretation is that substantive AI assistance is currently
more common in papers centered on relatively focused mathematical arguments,
whereas non-substantive AI use---such as language editing, coding, and
literature assistance---occurs across a broader range of manuscript types.
These comparisons, however, are purely descriptive and should not be given a
causal interpretation. The two groups may differ systematically in field,
topic, authorship, paper type, or other characteristics, so the results do not
imply that substantive AI use itself causes manuscripts to be shorter or to
contain fewer references.

%One possible interpretation is that substantive AI assistance is currently more common in manuscripts centered on relatively focused mathematical arguments, whereas confirmed-use manuscripts without substantive involvement span a broader range of manuscript types. This interpretation is only suggestive: the comparisons are descriptive and do not imply that substantive AI use causes manuscripts to be shorter or to cite fewer references, since the groups may differ systematically in field, topic, authorship, or other characteristics.

\section{Discussion and conclusion}\label{sec:discussion}

Our results show that AI use in mathematical research is increasingly rapidly. Although language and formatting remains the most
common individual use, a substantial fraction of confirmed-use manuscripts
report AI involvement in proof construction, formalization, problem
formulation, or other research tasks. The evidence therefore suggests that AI
is beginning to move beyond auxiliary assistance toward more direct
participation in mathematical research.
This shift is especially visible in work on open problems. We identify
hundreds of named open-problem records among substantive-use manuscripts. 
%including long-standing conjectures and problems from established collections
%such as the Erd\H{o}s Problems database. 
These records should not be interpreted
as independently verified AI solutions, but they show that AI is already being
used in research directed at unresolved mathematical questions.
Adoption is also highly uneven. Substantive AI use is concentrated in certain
mathematical fields, among a relatively small group of repeated users and
institutions, and within a few countries and model providers. 
%Such concentration may reflect differences in mathematical workflows, early-adopter
%communities, model access, and disclosure practices; our observational design
%does not distinguish among these explanations.
%
We also observe descriptive differences between substantive-use and
non-substantive-only manuscripts, most clearly in manuscript length. 
%These differences may reflect the types of mathematical work for which substantive AI
%assistance is currently most useful, but they should not be interpreted
%causally because the two groups may differ systematically in field, topic,
%authorship, and paper type.

Several limitations are important. First, our study measures \emph{disclosed}
AI use. Undisclosed use is not observable from our data, so the reported
prevalence should not be interpreted as the total prevalence of AI use in
mathematics. Second, our classifications are produced by an LLM-based review
of authors' descriptions and are therefore subject to classification error,
although our ongoing manual validation is designed to quantify this error.
Third, institution and country assignments rely partly on normalized and
inferred affiliation information and should be interpreted as broad
distributional summaries rather than precise rankings. Most importantly, our
classification of open-problem outcomes is not a substitute for peer review,
historical priority research, or independent verification of mathematical
correctness.

Despite these limitations, the evidence points to a rapidly changing research
landscape. The relevant question is no longer only whether AI systems
\emph{can} perform research-level mathematics, but increasingly where, how,
and by whom they are already being used to do so. The sharp growth observed
over only six months, together with the emergence of substantive use across
many fields and research problems, suggests that this transition is occurring
on a timescale that warrants continued empirical monitoring. Repeating this
disclosure audit over time, together with stronger human validation and
independent verification of mathematical claims, would help determine whether
the current ``gold rush'' represents a short-lived period of experimentation
or the beginning of a durable transformation in the practice of mathematical
research.

\section*{Disclosure of AI Use}

The authors used ChatGPT to assist with language editing, organization, and
manuscript preparation. Codex was used to assist with data collection, data
processing, data analysis, and figure generation. All AI-assisted outputs were reviewed by
the authors, who take full responsibility for the final manuscript.

\bibliographystyle{plainnat}
\bibliography{AI4Math}

\end{document}